\documentclass[manuscript,screen]{acmart}
\AtBeginDocument{%
  \providecommand\BibTeX{{%
    \normalfont B\kern-0.5em{\scshape i\kern-0.25em b}\kern-0.8em\TeX}}}

\setcopyright{acmcopyright}
\copyrightyear{2021}
\acmYear{2021}
\acmDOI{10.1145/1122445.1122456}

\acmConference[ACM CHI'21]{ACM CHI'21}{May 08--13, 2021}{Online, Earth}
\acmBooktitle{ACM CHI'21, June 03--05, 2021, Online, Earth}
\acmPrice{15.00}
\acmISBN{978-1-4503-XXXX-X/18/06}

\begin{document}

\title{A Way to a Universal VR Accessibility Toolkit}

\author{Felix J. Thiel}
\email{felix.thiel.18@ucl.ac.uk}
\orcid{0000-0002-7998-4270}
\author{Anthony Steed}
\email{A.Steed@ucl.ac.uk}
\orcid{0000-0001-9034-3020}
\affiliation{%
  \institution{Department of Computer Science, University College London}
  \streetaddress{66-72 Gower St}
  \city{London}
  \country{UK}
  \postcode{WC1E 6EA}
}

\renewcommand{\shortauthors}{Thiel and Steed, et al.}

\begin{abstract}
    \textbf{This work was presented at the ACM CHI 2021 Workshop on Design and Creation of Inclusive User Interactions Through Immersive Media.} \\ \\
  Virtual Reality (VR) has become more and more popular with dropping prices for systems and a growing number of users. However, the issue of accessibility in VR has been hardly addressed so far and no uniform approach or standard exists at this time. In this position paper, we propose a customisable toolkit implemented at the system-level and discuss the potential benefits of this approach and challenges that will need to be overcome for a successful implementation.
\end{abstract}

\ccsdesc[500]{Human-centered computing~Accessibility systems and tools}
\ccsdesc[500]{Human-centered computing~Virtual reality}

\keywords{Virtual Reality, Accessibility Systems and Tools}



\maketitle

\section{Introduction}
With the release of ever-cheaper virtual reality systems, the number of users of virtual reality (VR) is growing and its popularity is increasing. However, despite this trend, there is little work underway to address the unique accessibility challenges that come with VR. One of the main selling points of VR is that the user can interact with the virtual world in the same or at least a very similar way as they would in the real world. As a consequence of this, the user's body is primarily used as the input device for the applications which is a problem for physically impaired players. However, data gathered by other researchers indicates that this population is likely to be equally interested in the use of virtual reality as their able-bodied counterparts. In 2018, Beeston et al. \cite{beeston2018} tried to capture an image of the population of disabled players with a survey. They received 154 responses to their questionnaire and found that their participants do not differ much from able players in terms of play times and preferences. While this survey was focused solely on non-VR games, it still underpins the need for ways to make games, VR and non-VR alike, accessible to enable players of all abilities to share them. Based on this survey, it stands to reason that the rise in popularity of VR games will not be limited to the able-bodied population. Gerling et al. \cite{10.1145/3313831.3376265} also performed a survey on 25 wheelchair users about their views on virtual reality and found that there was general excitement about the potential of VR, though the participants also expressed concern about their ability to use it.

While this is a new problem for VR designers, accessibility issues have been widely explored in other media such as operating systems or televisions. Systems or devices in these classes are not designed to consider all kinds of impairments but they both provide tools to make them more accessible to users with specific impairments such as audio descriptions for deaf viewers or high contrast display modes for visually impaired computer users. However, for VR very little in terms of guidelines or tools exist that a seated player can use to make their experience more accessible to them. Some VR applications provide some accessibility options, but they are still in the minority. Examples of this are Half-Life: Alyx \cite{websiteHLAlyx} and Job Simulator \cite{websiteJobSimulator} that both offer subtitles and adjustments for seated players. However, these implementations are part of those games and can consequently not be used for other VR applications. Any other application remains inaccessible until their developer adds accessibility features.

We believe that this issue should not solely be left to individual developers but instead should be addressed on a system-level. It can not be expected that every developer adjusts their application to all potential impairments a player can have because they are manifold and it will most likely also not be an economically viable option. In the real world, the person's individual impairments are addressed by providing them with tools to overcome or at least alleviate the issue. These tools range from glasses to correct eye disorders, to powered wheelchairs and speech-generating devices. As the user's body is engaged with the input devices in VR, we find it fitting that a similar approach should be taken in VR that provides the user with a set of tools that they then can use and combine to fit their individual needs. They should also be able to take these tools, once configured, to other VR applications and use them in the same way one can use the same pair of glasses to read two different books.

\section{Proposal: A System-level Accessibility Toolkit} 
We propose that instead of expecting the developers of each individual game to provide accessibility tools, this problem should be approached on the level of the VR system. Currently, the communication between VR games and the hardware goes through one or multiple VR frameworks, often built by the hardware manufacturers. Examples are Oculus's OVR framework and Valve's SteamVR. Accessibility tools located on this layer would be in a prime location to intercept and influence data coming from and going to the VR hardware. In one direction, the game output going to the headset can be changed and augmented. In the other direction, the user's input can be altered or new emulated input fed in. This is analogous to tools such as the Magnifier that sits in the Windows operating system and manipulates the output image and scales the mouse input before passing it along.

The scope of these tools could range from simpler tools that focus on a single type of data (e.g. move the player's body, enhance audio, amplify haptics, etc.) to complex cross-modal tools that transfer information from one channel into another (e.g. auralising visual information). They could also be used to add additional controls or in an attempt to loosen the link between the control scheme and environment to allow for easier integration of additional input devices. The latter concept has been previously described as ``vehicle pattern'' \cite{Steed2019Vehicle}.

We also suggest that these accessibility tools are made in a modular fashion that allows the user to mix-and-match the individual components and switch them on and off depending on their needs.
This again can be found in examples from other media such as operating systems that allow the user to mix individual components such as the Magnifier, high-contrast interfaces, and text-to-speech.

\section{Related Work} 
A very similar approach was undertaken by Zhao et al. \cite{zhao2019seeingvr} with their SeeingVR toolkit. It was designed to make VR applications more accessible to people with low vision (i.e. not blind, but also not correctable by glasses). The outcome was a set of 14 tools that provide visual and audio augmentation for people with low vision. Similarly to our proposal, the individual tools can be selected, adjusted, and combined by the user to fit their specific needs. To make their tools compatible with a large number of VR games, they are injecting code into existing Unity applications. As a result, nine of their 14 tools can be used with any Unity application without changes to it. The remaining five tools require the developers to include an SDK that they developed into their game.
Their overall approach of accessibility tools that can be added on top of existing games is very similar to our proposal. However, they do it at the level of the game engine, which excludes VR games that are made with other game engines such as the Unreal Engine or the Source Engine and the approach is at risk to changes in the game engine rendering behaviour. We propose to have the accessibility tools operate on the level of the VR-system that is located between the hardware and the game engine.

We are currently working towards our proposed toolkit with our work on co-piloting in VR \cite{ThielSteed}. With co-piloting, the controls of the game are shared across multiple input devices and players. This allows two players to act as one so that the second player can help the VR player with actions they can not perform on their own. This is particularly useful with seated VR players that lack the mobility and reach of a standing player. In the past year, we have run experiments with this interaction technique and are currently on the way to develop this into a toolkit on its own. Similar to SeeingVR, our Co-Piloting prototypes can be used with any application that fulfils a certain requirement. While SeeingVR required it to be a Unity application, our prototypes just require the application to be based on the SteamVR framework, which includes a larger number of popular VR games. As such, our prototypes are already operating on the system-level and demonstrate the feasibility of accessibility tools on the system-level.

\section{Potential Benefits of a System-Level Approach}

\subsection{Increase in Accessibility}
The biggest benefit of a system-level approach is that it would increase the accessibility of any VR game, past and future, without any changes by the original developer. This does not only benefit the developers who have less effort in making their games accessible but also the players because they have a wider selection of games that are accessible to them.

\subsection{Customisation and Ease of Use}
Another benefit is the potential of customisation. If the toolkit is based on modules that address different impairments, the players can use these building blocks to create a suite that is tailored to their needs. These combinations could also be saved as presets to remove the need to configure every game. It would also allow multiple users to share a system without the need for lengthy reconfiguration in between.

\subsection{Uniform Access}
A single toolkit that is usable across all VR applications also has the benefit that it will provide uniform access to the accessibility tools. When each game developer develops their own tools, their controls, capabilities, and availability  will likely differ between games and make using them more complex. A shared toolkit at the system-level could be used with all games and provide the same user experience and controls in all of them.

\section{Challenges of a System-Level Approach}

\subsection{System Integration and Support}
One big challenge is a potential lack of support by the system manufacturers. One issue we are struggling with during the development of our tools is that we need to use library hooks and undocumented interfaces to manipulate the player's virtual body reliably because the required matrices are not exposed to the outside. This also requires us to use an outdated version of SteamVR because later updates are not compatible with the library hooks any more. To prevent this kind of difficulties, a system-level accessibility toolkit will require support from the system manufacturers or upcoming standards like OpenXR. This support could be realised through an Accessibility API or an SDK that exposes the internals state of the VR systems.

\subsection{System Abuse}
To achieve their full potential, the tools will require extensive access to the input and output of the VR system. This comes with the danger that they get abused for cheating. A tool that moves the position of the virtual body in the VR scene can help a player with mobility issues play, but it can also give additional mobility and an unfair advantage in competitive games. This is not an easy challenge. The game could be informed about the player using any accessibility tools because these tools will use functionality and APIs offered by the VR system, but excluding those players or putting them on a separate ranking will once again separate players with and without impairments.

\subsection{Development and Maintenance}
Another challenge is the question of who will provide and support these tools. The accessibility tools of TV and operating systems are developed and maintained by the system manufacturers which could promise a good integration into the system and reliable maintenance. However, the manufacturers might just cover the most frequent impairments which limits the usefulness of the full system. A community-driven toolkit might offer a larger variety of tools that cover more impairments but brings the risk of creators and maintainers to later abandon their tools. This less structured approach also bears the risk of inconsistent control schemes between tools that interfere with each other.

\subsection{Control Conflicts}
Another challenge of this approach is that the controls of any of the system-level tools may conflict with the already existing controls of the games. This is not an issue for accessibility tools built into games, because the developers can incorporate them into their control scheme. One potential solution for this could be to use input channels that have not been used much as game input so far such as voice input or gaze input. Another possible solution for this could be a less rigid approach to control schemes as currently used with SteamVR. SteamVR provides the user with a system in which they can remap the game functions on their input device and share their mappings with the community. This would make the creation and distribution of mappings that are compatible with the controls of the accessibility tools much easier and more user-friendly.



\begin{acks}
This project has received funding from the European Union’s Horizon 2020 research and innovation programme under grant agreement N° 856998.

\end{acks}

\bibliographystyle{ACM-Reference-Format}
\bibliography{sample-base}



\end{document}